\documentclass[]{aa}
\usepackage{graphicx}
\usepackage{color}
\usepackage{txfonts}

\usepackage{amstext}

\begin{document}

\title{Formation and evolution of coronal rain observed by SDO/AIA on February 22, 2012 }

\author{Vashalomidze, Z.\inst{1,2}, Kukhianidze, V.\inst{2}, Zaqarashvili, T.V. \inst{1,2}, Oliver, R.\inst{3}, Shergelashvili B.\inst{1,2}, Ramishvili G.\inst{2}, Poedts
S.\inst{4} and De Causmaecker P.\inst{5} }

\institute{Space Research Institute, Austrian Academy of Sciences, Schmiedlstrasse 6, 8042 Graz, Austria \\
             \email{[teimuraz.zaqarashvili]@oeaw.ac.at}
                                           \and
             Abastumani Astrophysical Observatory at Ilia State University, Cholokashvili Ave.3/5, Tbilisi, Georgia\\
                                                       \and
            Departament de F\'isica, Universitat de les Illes Balears, E-07122, Palma de Mallorca, Spain\\
                                                      \and
        Dept. of Mathematics, Centre for Mathematical Plasma Astrophysics, Celestijnenlaan 200B, 3001, Leuven, Belgium\\
                                                          \and
            KU Leuven Campus Kortrijk, E. Sabbelaan 53, 8500 Kortrijk, Belgium
            }

\date{Received / Accepted}

\abstract{The formation and dynamics of coronal rain are currently
not
fully understood. Coronal rain is the fall of cool and dense blobs formed by thermal instability in the solar corona towards the solar surface with acceleration smaller than gravitational free fall.}{We aim to study the observational evidence of the formation of coronal rain and to trace the detailed dynamics of individual blobs.}{We used time series of the 171 \AA\, and 304 \AA\, spectral lines obtained by the Atmospheric Imaging Assembly (AIA) on board the Solar Dynamic Observatory (SDO) above active region AR 11420 on February 22, 2012.}{Observations show that a coronal loop disappeared in the 171 \AA\ channel and appeared in the 304 \AA\ line$\text{}\text{}$ more than one hour later, which indicates a rapid cooling of the coronal loop from 1 MK to 0.05 MK. An energy estimation shows that the radiation is
higher than the heat input, which indicates so-called catastrophic cooling. The cooling was accompanied by the formation of coronal rain in the form of falling cold plasma. We studied two different sequences of falling blobs. The first sequence includes three different blobs. The mean velocities of the blobs were estimated to be 50 km s$^{-1}$, 60 km s$^{-1}$ and 40 km s$^{-1}$. A polynomial fit shows the different values of the acceleration for different blobs, which are lower than free-fall in the solar corona. The first and second blob move along the same path, but with and without acceleration, respectively. We performed simple numerical simulations for two consecutive blobs, which show that the second blob moves in a medium that is modified by the passage of the first blob. Therefore, the second blob has a relatively high speed and no acceleration, as is shown by observations. The second sequence includes two different blobs with mean velocities of 100 km s$^{-1}$ and 90 km s$^{-1}$, respectively.}{The formation of coronal rain blobs is connected with the process of catastrophic cooling. The different acceleration of different coronal rain blobs might be due to the different values in the density ratio of blob to corona. All blobs leave trails, which might be a result of continuous cooling in their tails.}

\keywords{Sun: corona -- Sun: coronal rain}

\authorrunning{Vashalomidze et al.}

\titlerunning{Formation and evolution of coronal rain}

\maketitle

\section{Introduction}

Observations show rapidly formed cool, dense plasma blobs falling down towards the surface in the hotter solar corona. This phenomenon is known as coronal rain. Coronal rain is observed to occur in active region coronal loops, where the
blobs are formed by thermal instability and then fall towards their footpoints. Coronal rain is also related to solar prominences, especially to cool blobs that detach from the main body of the prominence and fall down towards the photosphere.

Schrijver (2001) detected clumps of relatively cool plasma moving at speeds of 100 km s$^{-1}$ along active region coronal loops. The downward acceleration of the clumps was no more than 80 m s$^{-2}$, which is much lower than the free-fall acceleration by surface gravity. De Groof et al. (2004) detected a similar phenomenon with SOHO/EIT in 304 \AA\ line. Then, De Groof et al. (2005) analysed the coronal rain blobs simultaneously with 304 \AA\ images from SOHO/EIT and H$\alpha$ images from Big Bear Observatory. They detected blobs with speeds of between 30 and 120 km s$^{-1}$ and with accelerations much lower than the gravitational free fall. Zhang \& Li (2009) analysed 26 coronal rain events in Ca II filtergrams using images of the Solar Optical
Telescope (SOT) on board Hinode and found two types of blobs:  fast and slow, with average speeds of 72 km s$^{-1}$ and 37 km s$^{-1}$, respectively. Antolin at al. (2010) analysed SOT/Hinode Ca II H line data and found that the blobs started to fall down from the height of 60-100 Mm with low ($\sim$ 30-40 km s$^{-1}$) speed, but accelerated to high ($\sim$ 80-120 km s$^{-1}$) speed in the lower parts of loops. The accelerations were found to be on average substantially lower than the solar gravity component along the loops. Detailed statistical studies of coronal rain events using H$\alpha$ data from the CRisp Imaging Spectro Polarimeter (CRISP) instrument at the Swedish Solar Telescope (Antolin \& Rouppe van der Voort 2012, Antolin et al. 2012) showed that the kinematics and morphology of on-disk and off-limb blobs are basically the same. These authors combined the apparent motion and Doppler velocity of the blobs and estimated the average velocity and acceleration to be 70 km s$^{-1}$ and 83.5 m s$^{-2}$, respectively.

\begin{figure*}
\begin{center}
\includegraphics[width=15cm]{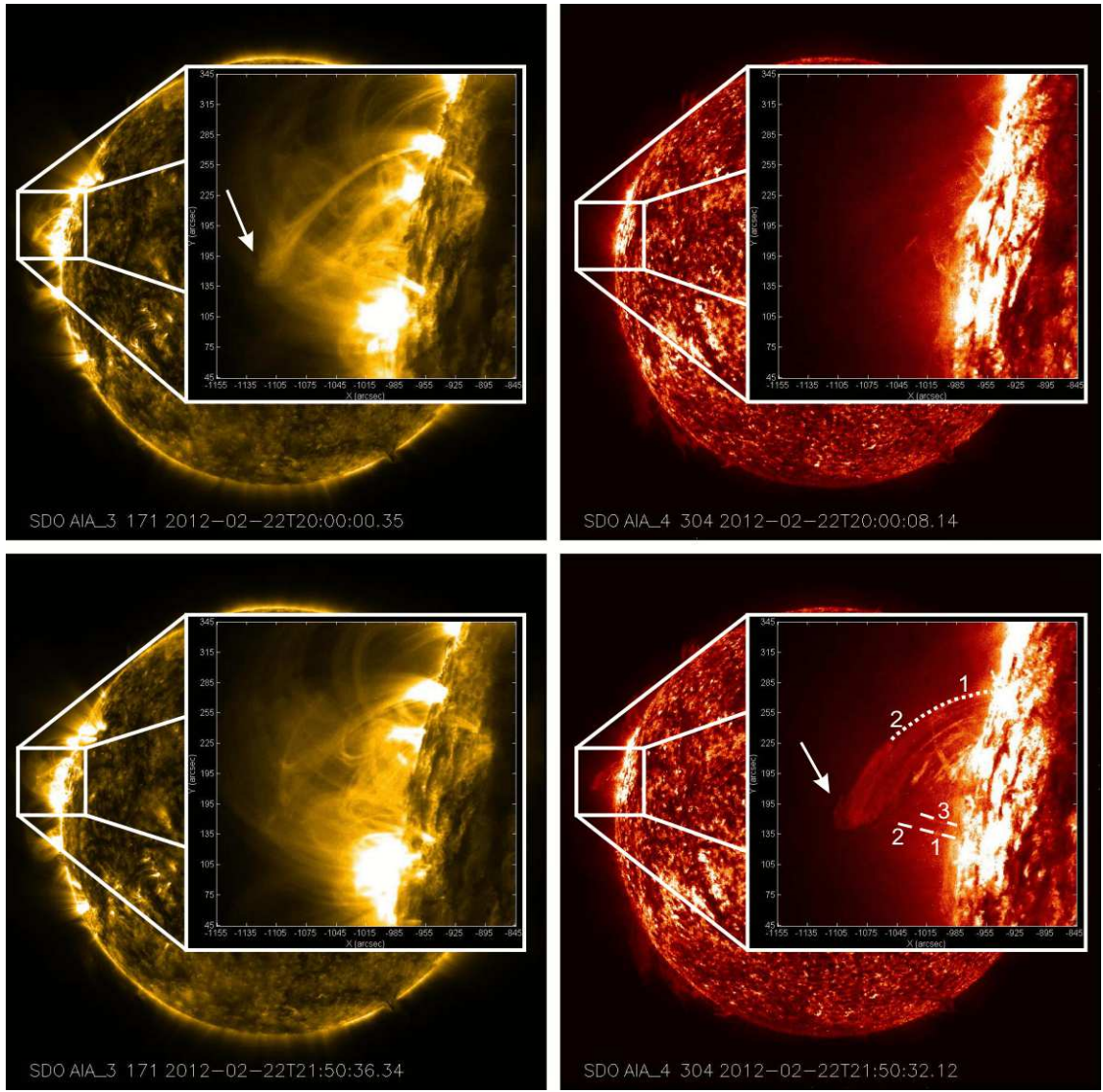}
\end{center}
\caption{Cooling of a coronal loop as it is seen in coronal images above active region AR 11420 on February 22, 2012. Upper panels: images in the 171 \AA\, (left) and 304 \AA\, (right) lines at UT 20:00. Lower panels: images in the 171 \AA\, (left) and 304 \AA\, (right) lines after 2 hours, i.e., at UT 21:50. The large boxes show zoom-ins of the active region corona. The dashed white lines in the lower right panel show the trajectories of the coronal rain blobs for event 1: the lower line corresponds to the trajectories of the first and second blob, while the upper line corresponds to the trajectory of the third blob. The dotted white line shows the trajectory of the coronal rain blobs for event 2: first and second blobs move on same path. The corresponding movies in the 171 \AA\,  and 304 \AA\, lines are available in the online journal.}
\end{figure*}

\begin{figure*}
\begin{center}
\includegraphics[width=\columnwidth]{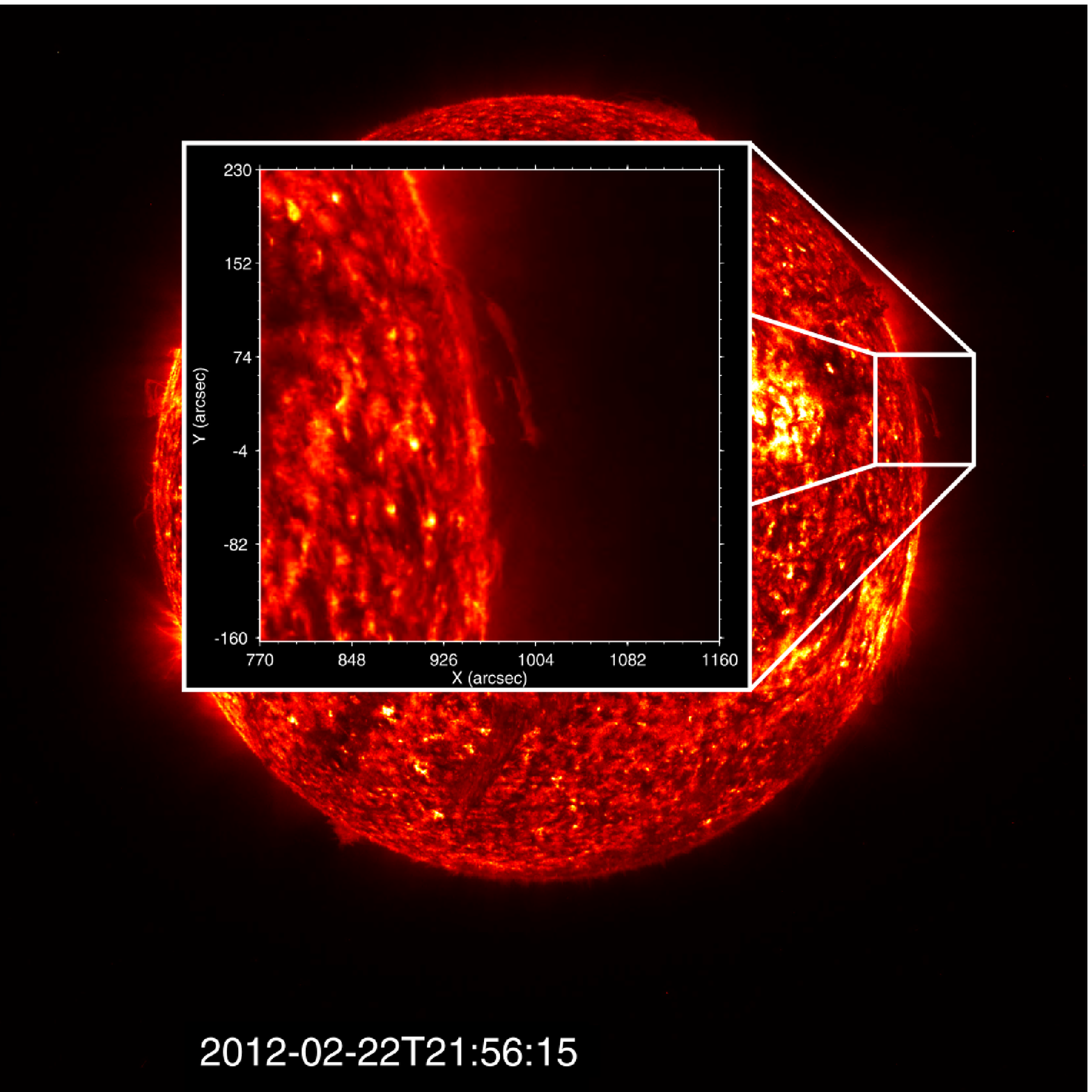}
\end{center}
\caption{Coronal loop image in 304 \AA\ taken by STEREO B above active region AR 11420 at UT 21:56 on February 22, 2012. The large box shows a zoom-in of the active region corona.}
\end{figure*}

Liu at al. (2012) studied the formation of a prominence with images taken in 193 \AA\,, 171 \AA\ and 304 \AA\ by the Atmospheric Imaging Assembly (AIA) on board the Solar Dynamic Observatory (SDO). They concluded that the prominence mass is not static but maintained by condensation at a high
estimated rate against a comparable drainage through numerous vertical downflow threads. These downflows occurred in the form of cool mass blobs. The cool, dense blobs that resemble the coronal rain started to fall down from a height of 20-40 Mm. The velocity of the blobs had a narrow Gaussian distribution with a mean of 30 km s$^{-1}$, while the downward acceleration distribution had an exponential drop with a mean of 46 m s$^{-2}$.

The plasma condensation is caused by thermal instability, which leads to the formation of coronal rain through catastrophic cooling whenever radiative losses locally overcome the heating input (Parker 1953, Field 1965, Antiochos et al. 1999, Schrijver 2001). Numerical simulations by M\"uller et al. (2003, 2004, 2005) also indicated that the thermal instability can be responsible for the formation of cold condensation and consequently the coronal rain. On the other hand, Murawski et al. (2011) suggested that null points in the corona are locations where preferential cooling can occur through thermal mode (entropy mode) generation. This mode, which is characterised by a local density enhancement and a decrease in temperature, is triggered by a local loss of pressure in null points. Consequently, cold blobs are formed and can be carried away from the null points by reconnection outflows that
form the coronal rain events.

\begin{figure*}[t]
\vspace*{1mm}
\begin{center}
\includegraphics[width=12cm]{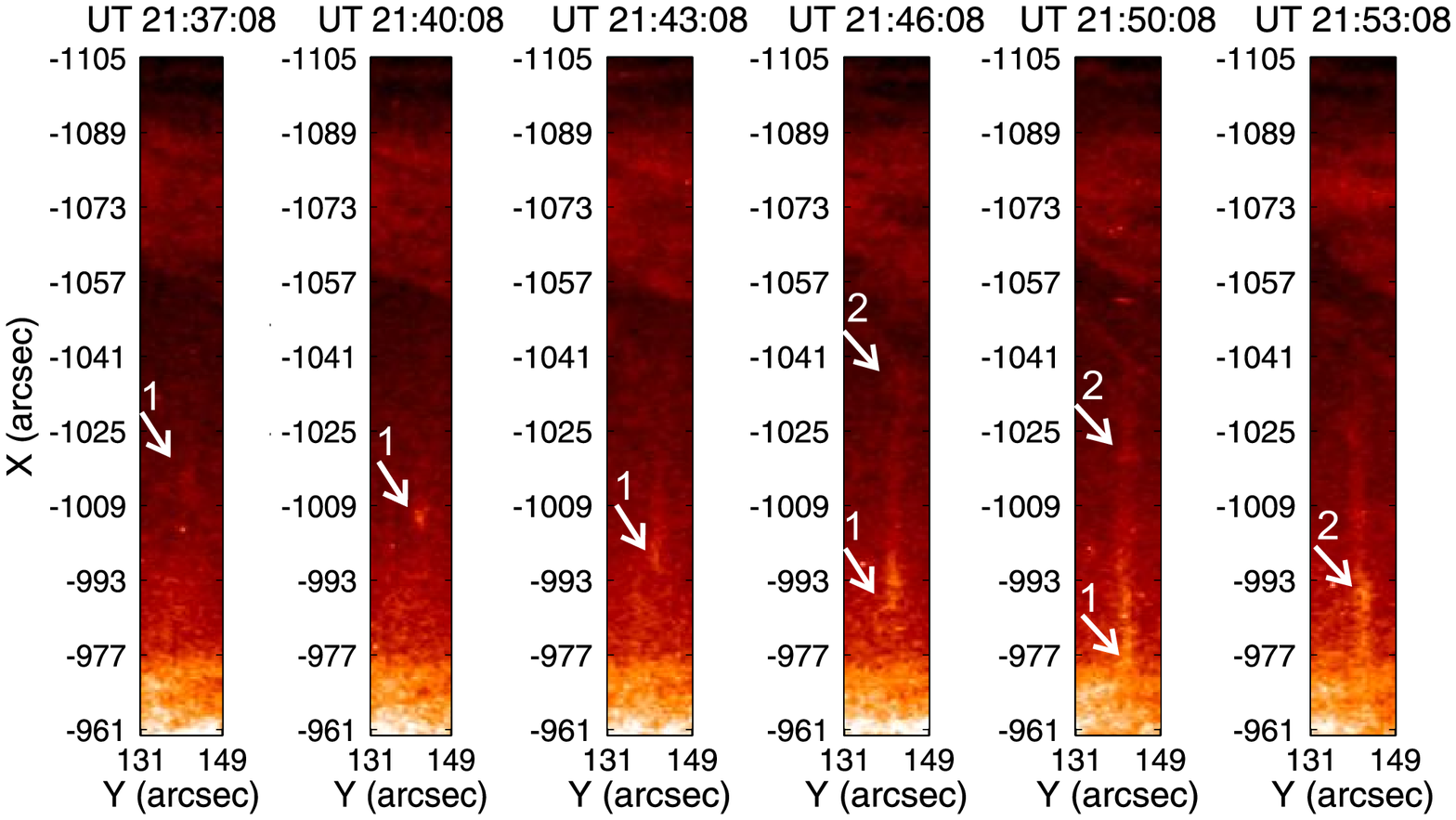}
\includegraphics[width=10cm]{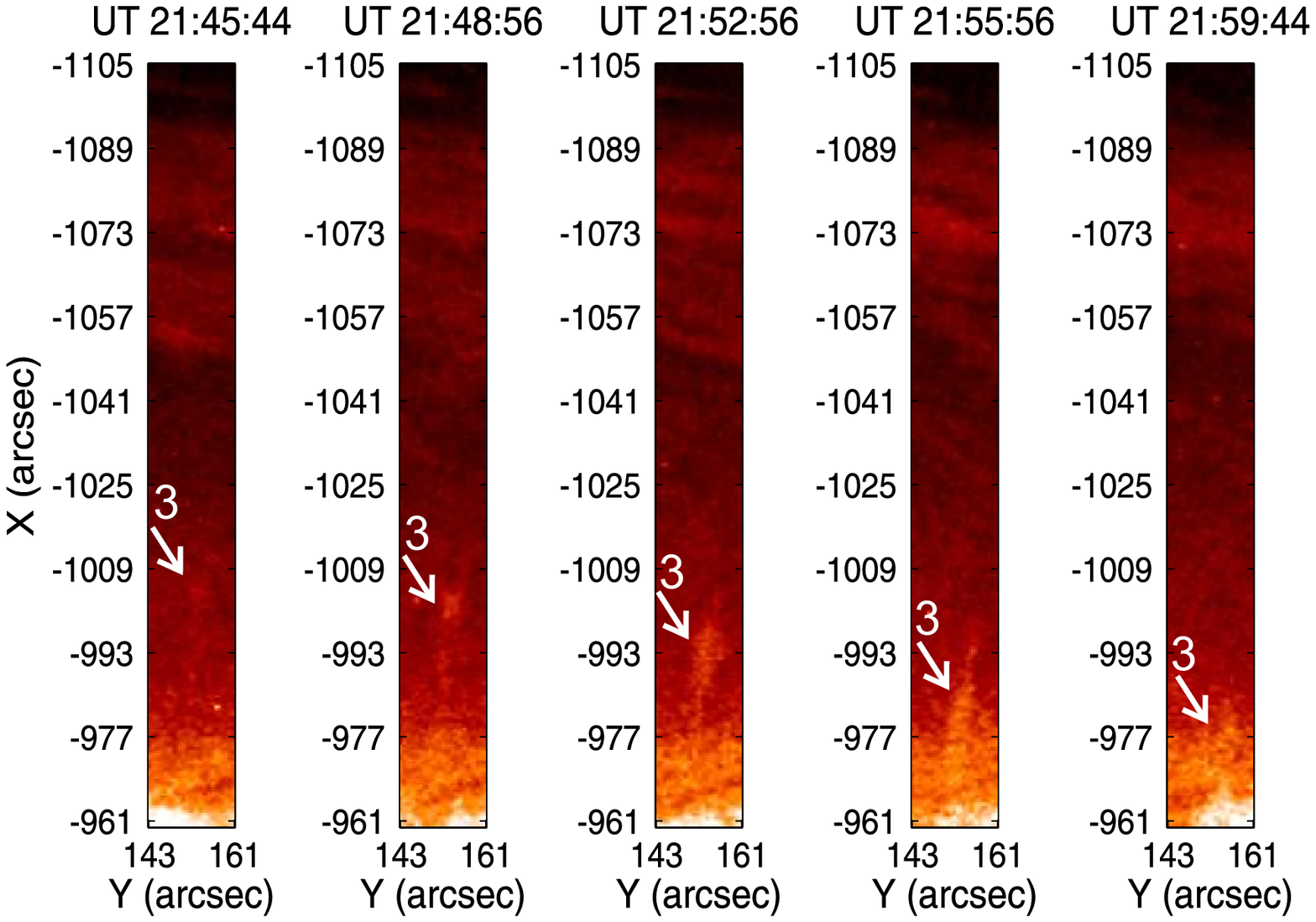}
\end{center}
\caption{Six consecutive images of the first and second blob (upper panel) and five consecutive images of the third blob (lower panel) from the first sequence in the 304 \AA\ spectral line. The size of each box is 18x145 arc sec and it coincides at the dashed white line in the lower right panel in Fig. 1. The blobs are indicated by white arrows in consecutive images.}
\end{figure*}

Coronal rain blobs are falling with much lower acceleration than the gravitational free-fall acceleration in active region loops. Simulations have shown that the observed dynamics of blobs might be explained by effects of gas and magnetic pressure (Mackay and Galsgaard 2001, M\"uller et al. 2003; Antolin et al.
2010). Antolin and Verwichte (2011) suggested that the pressure from transverse waves in the coronal loops might be
responsible for the observed low downward acceleration of coronal rain. Recently, Oliver et al. (2014) found that heavy condensation gives rise to a dynamical rearrangement of the coronal pressure that results in the formation of a large pressure gradient that opposes gravity. Consequently, this pressure gradient may force the coronal rain blobs to fall with lower acceleration.

We used SDO/AIA observations to study the formation and dynamics of coronal rain over active region AR 11420 on 22 February, 2012. We detected rapid cooling of a coronal loop from coronal to transition region temperature. This triggered the appearance of cool blobs that fell down along magnetic field lines. We studied the detailed dynamics of two sequences of several blobs including velocity, acceleration, and forming height.

\section{Observations}

We used the observational data obtained by AIA on board the SDO (Pesnell et al. 2011) on February 22, 2012. AIA provides full-disk observations of the Sun in three ultra-violet (UV) continua at 1600 \AA\,, 1700 \AA\,, 4500 \AA,\, and seven extreme ultra-violet (EUV) narrow bands at 171 \AA\,, 193 \AA\,, 211 \AA\,, 94 \AA\,, 304 \AA\,, 335 \AA\,, and 131 \AA\, with 1.0'' resolution and 12 s cadence (Lemen at al. 2011). We used time-series of the 171 \AA\, and 304 \AA\, wavelengths. The 171 \AA\, wavelength is dominated by the quiet corona and upper transition-region Fe IX line, corresponding to a temperature of $\sim$ $10^{5.8}\ \mathrm{K,}$ and the 304 \AA\, wavelength is dominated by the transition-region line of He II, corresponding to a temperature of $\sim$ $10^{4.7}\ \mathrm{K}$. The time-series of SDO/AIA data was reduced by the standard SSW cutout service.

We also used data from Extreme Ultra-Violet Imager (EUVI) narrow bands at 171 \AA\ and 304 \AA\, of the Solar Terrestrial Relations Observatory (STEREO) on February 22, 2012.

\section{Formation and dynamics of coronal rain blobs}

Coronal rain was observed in the 304 \AA\, line above active region AR 11420 on 22 February, 2012. The formation of the coronal rain was connected to a coronal loop in the overlying corona. Figure 1 shows the consecutive images in the 171 \AA\, and 304 \AA\, lines with a time interval of  $ \text{}\text{about two}$  hours. The coronal loop is seen in the 171 \AA\, line at UT 20:00 (shown by the white arrow in Fig. 1, upper left panel), but it is absent from the 304 \AA\, image (Fig. 1, upper right panel). However, during the next one to two hours, the coronal loop permanently became faint in 171 \AA\, and appeared in the 304 \AA\, line. At UT 21:50, the coronal loop is already visible in 304 \AA,\, but not in 171 \AA\, (Fig. 1, lower panels). Because the 171 \AA\, line corresponds to hotter plasma than the 304 \AA\, line, it is clear that the coronal loop has cooled from $\sim$ $10^{5.8}$ to $\sim$ $10^{4.7}$ $\mathrm{K}$. The coronal loop (or loop system) has slightly different width in the two spectral lines and reaches $\sim$ 30 Mm in 304 \AA\, near the apex, which is located at $\sim$ 70 Mm above the surface.

The coronal loop seems to be located in the coronal loop system.
This is significantly bent with its plane directed roughly along the line of sight. However, this might be only projection effects and loops may have other directions. To understand the 3D structure of coronal loop system, we tried to use the data from the Solar Terrestrial Relations Observatory (STEREO). Unfortunately, there was only one image in the 171 \AA\ line during the observed period, and it could not help us to understand the real structure of the active region. On the other hand, the coronal loops are visible in the 304 \AA\ line. Figure 2 shows the image taken by STEREO B in the 304 \AA\ line at UT 21:56. We were unable to resolve the coronal rain blobs because of the limited spatial resolution of STEREO. But the feature that is observed above the limb in Fig. 2 corresponds well to the geometry that is observed with AIA.

When the coronal loop cools down, bright blobs start to appear in the 304 \AA\ line in SDO data. We observed two different sequences of coronal rain. The first sequence appears below the coronal loop and the blobs fall down almost vertically. The second sequence appears near the apex of the coronal loop, and the coronal rain material moves along the loop on an inclined trajectory, which is clearly seen in the 304 \AA\ line (see on-line movie). We consider the two sequences of coronal rain separately.

\subsection{First sequence of blobs}
The first blob of this sequence appears at UT 21:37 at a height of $\sim$ $ 4 \times 10^4 $ km.  Figure 3 (upper panel) shows six consecutive images of the blob with intervals of $\sim$ 3 min. After the appearance, the blob starts to fall down along a slightly inclined trajectory, probably along a coronal loop (see the lower dashed line in the lower right panel of Fig. 1). The width and length of the blob at its first appearance are $\sim$ 1400 km and $\sim$ 3600 km, respectively. The width of the blob remains almost the same, but the length increases during the fall. Interestingly, the blob leaves a trail (see last four images in the upper panel of Fig. 3), which might be a result of continuous cooling in the blob tails (Fang et al. 2013). The blob reaches the surface $\sim$ 13 min after its first appearance with a mean speed of $\sim$ 50 km s$^{-1}$.

The second blob appears at UT 21:46 at a height of $\sim$ $6 \times 10^4$ km. This blob follows the same trajectory as the first blob and reaches the surface $\sim$ 17 min after its first appearance with a mean speed of $\sim$ 60 km s$^{-1}$ (see the last three images in the upper panel of Fig. 3).

The third blob appears simultaneously with the second blob, that
is, at UT 21:50, at a height of $\sim$ $3 \times 10^4$ km, but possibly moves along a different loop (see the upper dashed line in the lower right panel of Fig. 1). The third blob reaches the surface in $\sim$ 12 min with a mean speed of $\sim$ 40 km s$^{-1}$ (see the five consecutive images in the lower panel of Fig. 3).

Figure 4 shows the $x$-$t$ cuts ($x$ denotes the vertical direction) during the time interval UT 21:30-22:15 in the 304 \AA\ spectral line. The width of each cut is 1 pixel, which approximately equals 0.6 arc sec, and the length is 181 pixels. The trajectories of the three blobs are indicated by white arrows. The plots show
that the trajectories of the first and third blobs are more parabolic, while the trajectory of the second blob is almost linear.

We next obtained the positions of the blobs at different times during the fall. Figure 5 shows the plots of height vs time, the first (red points) and the third (blue points) blobs. These points constitute the trajectories of the blobs during the fall. Then we fitted the points with a quadratic polynomial $h=h_0-V_0 t-a t^2/2$, where $h_0$ is the initial height of the blob, $h$ is the height, $t$ is the time, $V_0$ is the initial velocity, and $a$ is the acceleration. The fittings were performed with and without initial velocity. The estimated acceleration without initial velocity (i.e. for $V_0=0$) is 120 m s$^{-2}$ for the first blob and 136 m s$^{-2}$ for the third blob. On the other hand, the fitting with a non-zero initial velocity yields the following parameters: $V_0$=12 km s$^{-1}$ and $a$=92 m s$^{-2}$ for the first blob and $V_0$=22 km s$^{-1}$ and $a$=74 m s$^{-2}$ for the third blob. The accelerations of both blobs are lower than the gravitational acceleration in both cases (i.e. with and without initial velocity). We did not plot the trajectory of the second blob because it is almost linear with zero acceleration and a mean speed of $\sim$ 60 km s$^{-1}$.

The polynomial fit shows that the initial velocity reduces the estimated accelerations of the first and third blob, which correspond to the values found in previous works (Schrijver 2001, Antolin \& Rouppe van der Voort 2012). But the acceleration of the second blob, which follows the same path (or the same loop) as the first blob, is almost zero. To try to understand the different behaviour of the two blobs that fall along the same magnetic flux tube, we performed a few numerical simulations based on the work of Oliver et al. (2014). We considered a vertically stratified magnetic tube whose base density and scale-height are those of Oliver et al. (2014); more details are given in that work. At $t=0,$ a fully formed blob with zero initial velocity is placed at a height of 40~Mm inside the magnetic tube; this condensation
corresponds to the first blob in our observations. To reproduce its height versus time variation, shown in Figs.~4 and 5, we need to adjust the initial blob density. The value $6\times 10^{-13}$~g~cm$^{-3}$ yields a good agreement between the observed path and the results of this numerical simulation (see the squares and solid line of Fig.~6). Our numerical simulations also take into account the second blob by including a mass injection at a height of 60~Mm at which the second blob is formed at $t=630$~s. This time corresponds to the delay between the appearance of the two blobs in our observations. The initial density of the second blob also needs to be adjusted so that its dynamics is similar to the observed one. The value $10^{-14}$~g~cm$^{-3}$ gives a reasonable agreement between the observations and the numerical simulations (see the triangles and dashed line in Fig.~6). It is clear that in this numerical simulation the second blob very rapidly achieves a roughly constant speed, as observed. The reason for this fast acceleration is that by the passage of the first blob, all the mass in the magnetic tube below and above it is set into a falling motion. Hence, the second blob forms in a moving plasma that imparts a downward push to the blob. In this simulation, however, the second blob has a much lower density than the first blob, and so one would expect a very different brightness in the 304~$\AA$ line. Figure~3 shows that this is not the case, and the reason surely lies in the simplicity of our numerical simulations. Likewise, the plasma might be rarefied after the passage of the first blob, so that the second encounters less friction, which leads to a
higher speed of the second blob (M\"uller et al. 2003). Recent numerical simulations of Fang et al. (2013) showed that a second blob has a lower acceleration than the first one in some cases. Therefore, this phenomenon may depend on the particular conditions in coronal loops.

\subsection{Second sequence of blobs}

In addition to these blobs, there is a more powerful coronal rain event after the cooling of the coronal loop: cool plasma starts to move along an inclined trajectory, probably along an inclined magnetic field of the coronal loop system (see movie in the 304~$\AA$ line, which is available in the online journal).
This is a flow-like event of coronal rain, which might consist of many smaller blobs, but we could not see them owing to the resolution limit. We were only able to identify two different blobs. The first blob appears at UT 21:41 at a height of $\sim$ $9.5 \times 10^4 $ km and moves along an inclined trajectory (see the dotted line in the lower right panel of Figure 1). The width and the length of the blob at its first appearance are $\sim$ 1400 km and $\sim$ 2200 km, respectively. The width of the blob remains almost the same, but the length increases during the fall. The blob reaches the surface $\sim$ 13 min after the first appearance with a mean speed of $\sim$ 100 km s$^{-1}$. The second blob appears at UT 21:57 at a height of $\sim$ $8.6 \times 10^4$ km. The blob follows the same trajectory as the first blob and reaches the surface $\sim$ 13 min after the first appearance with a mean speed of $\sim$ 90 km s$^{-1}$.

\begin{figure}[t]
\vspace*{1mm}
\begin{center}
\includegraphics[width=\columnwidth]{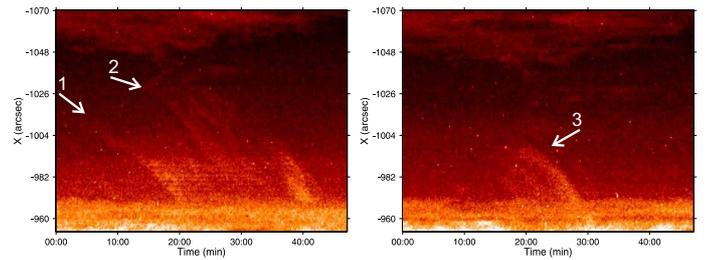}
\end{center}
\caption{$x$-$t$ cuts show the trajectories of the blobs from the first sequence during UT 21:30-22:15 in the 304 \AA\ line. The left plot corresponds to the cut along the path of the first and second blob (indicated by the lower dashed line in the lower right panel of Fig. 1). The right plot corresponds to the cut along the path of the third blob (indicated by the upper dashed line in the lower right panel of Fig. 1).}
\end{figure}

\begin{figure}[t]
\vspace*{1mm}
\begin{center}
\includegraphics[width=8cm]{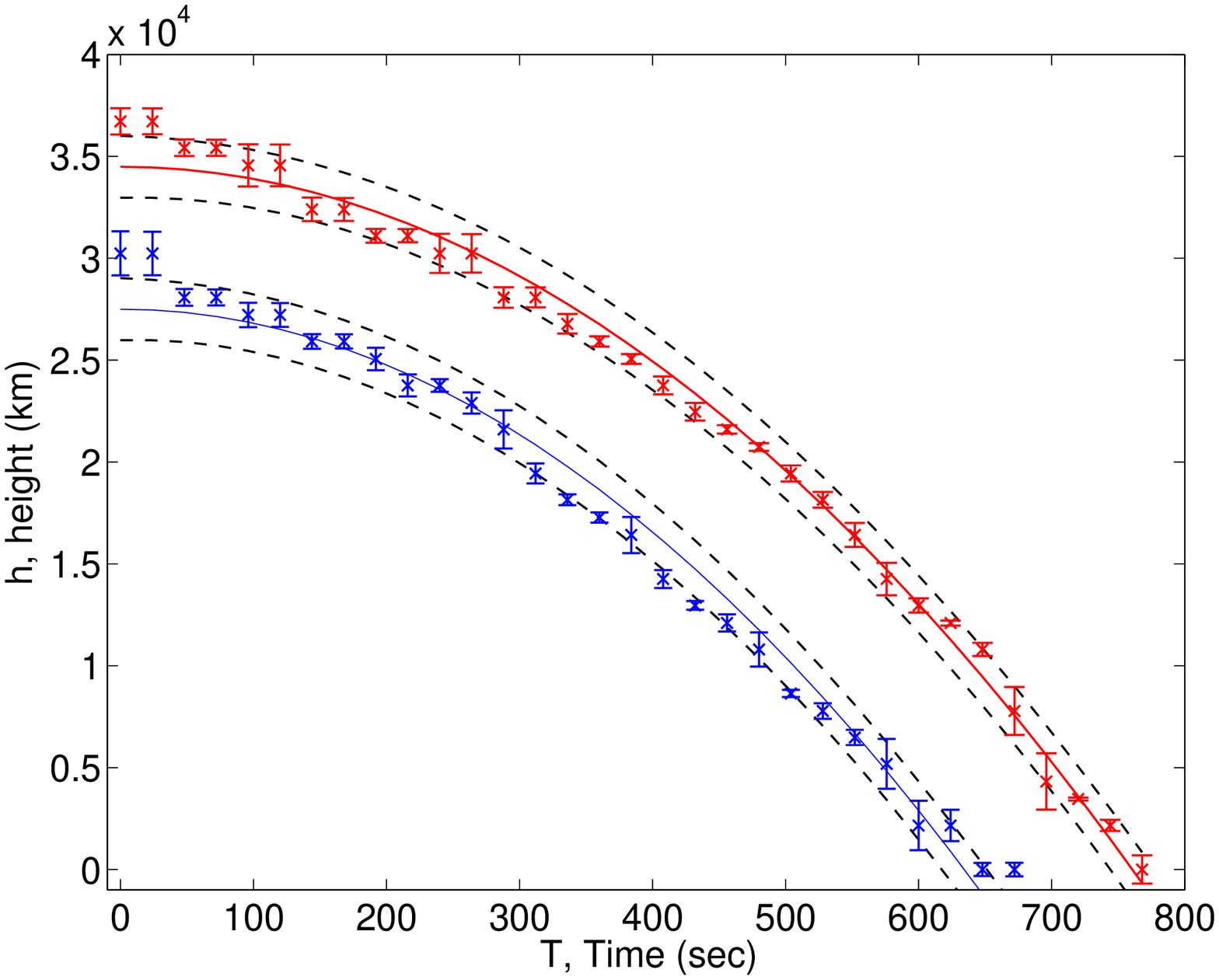}
\includegraphics[width=8cm]{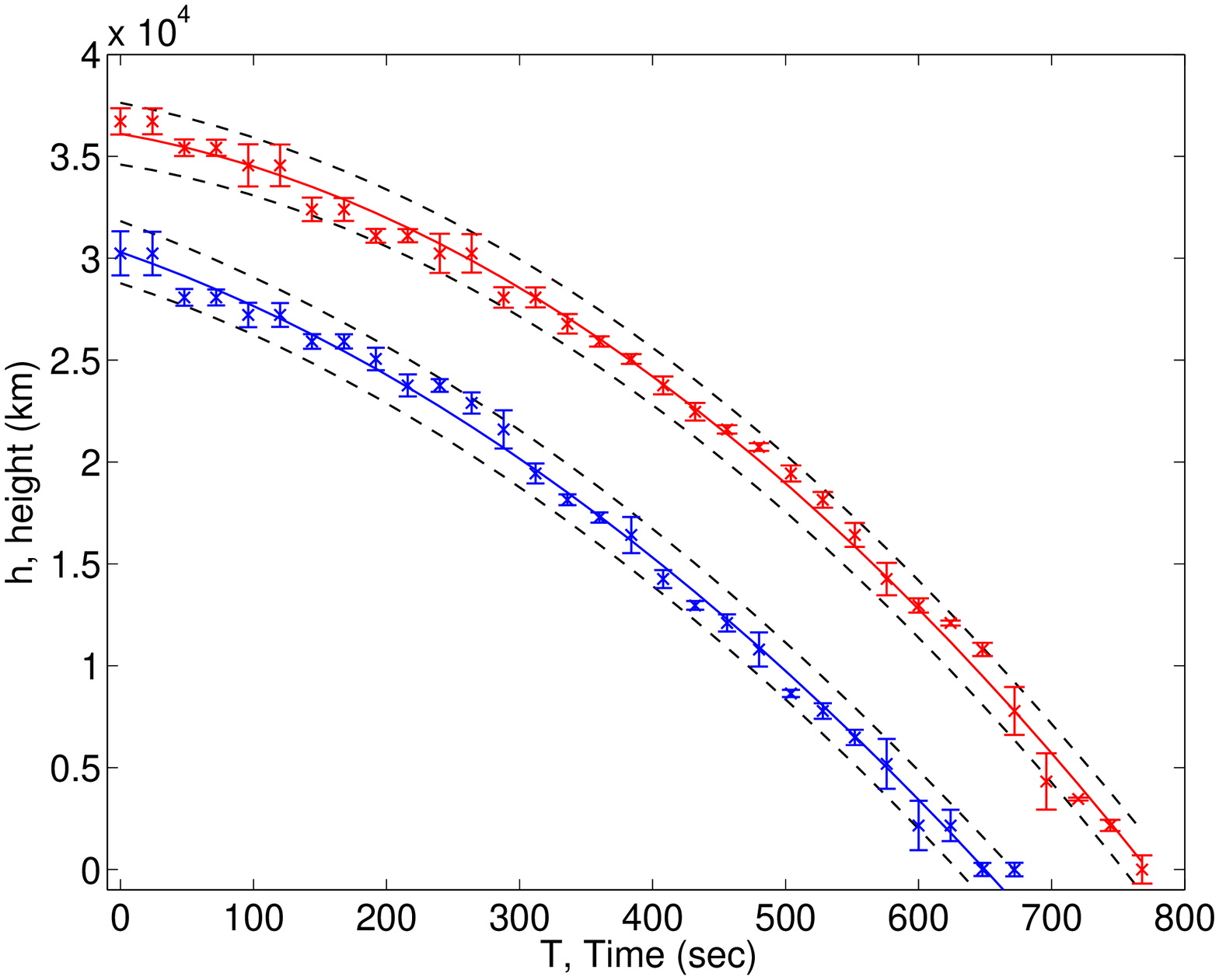}
\end{center}
\caption{Height vs time for the first (red dots) and the third (blue dots) blob from the first sequence. Fitted curves are the trajectories with constant acceleration for the first (red curve) and third (blue curve) blob. Upper panel: fitted trajectories with zero initial velocity. Lower panel: fitted trajectories with nonzero initial velocity. Error bars show the standard deviations, while the dashed curves show 95 \% confidence levels for the polynomial fitting.}
\end{figure}

\begin{figure}[t]
\vspace*{1mm}
\begin{center}
\includegraphics[width=6cm, angle=270]{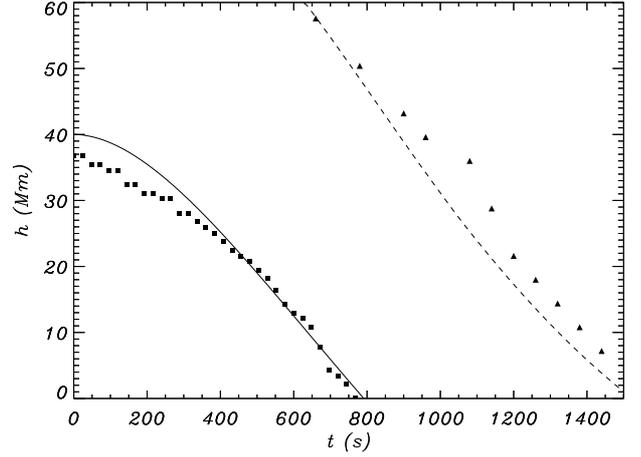}
\end{center}
\caption{Observed height versus time of the first (squares) and second blobs (triangles) from the first sequence. The solid and dashed lines show the dynamics of the two blobs obtained with the numerical simulation described in the text.}
\end{figure}

Figure 7 shows the plots of distance vs time for the first (red points) and the second (blue points) blob of the second sequence. The estimated acceleration along the path for the first blob with an initial velocity of $V_0$=80 km s$^{-1}$ is $a$= $\sim$130 m s$^{-2}$. Note that this is the effective acceleration because the blob moves along an inclined trajectory. The trajectory is curved with regard to the vertical, but we can estimate a mean angle of $\sim$ 45 degree, then the acceleration is lower than the effective free-fall acceleration of the solar atmosphere. On the other hand, the acceleration for the second blob along the inclined trajectory with an initial velocity of $V_0$=60 km s$^{-1}$ is $a$=100 m s$^{-2}$, which is lower than that of the first blob.

The second sequence consists of two blobs that fall along the same path, similar to blob 1 and 2 of the first sequence. The difference between these two sequences is that the second blob has zero acceleration in the first sequence and non-zero acceleration in the second one. The reason for this different dynamics probably is that blob 2 in the second sequence appears after blob 1 has reached the lowest part of its trajectory, while blob 2 in the first sequence appears when blob 1 still falls down. Then the two blobs of the first sequence are connected dynamically, but the two blobs of the second sequence are not. This fact again confirms the results of our numerical simulation: the motion of the first blob influences the dynamics of the second blob when they fall along the same path.

\begin{figure}[t]
\vspace*{1mm}
\begin{center}
\includegraphics[width=7cm]{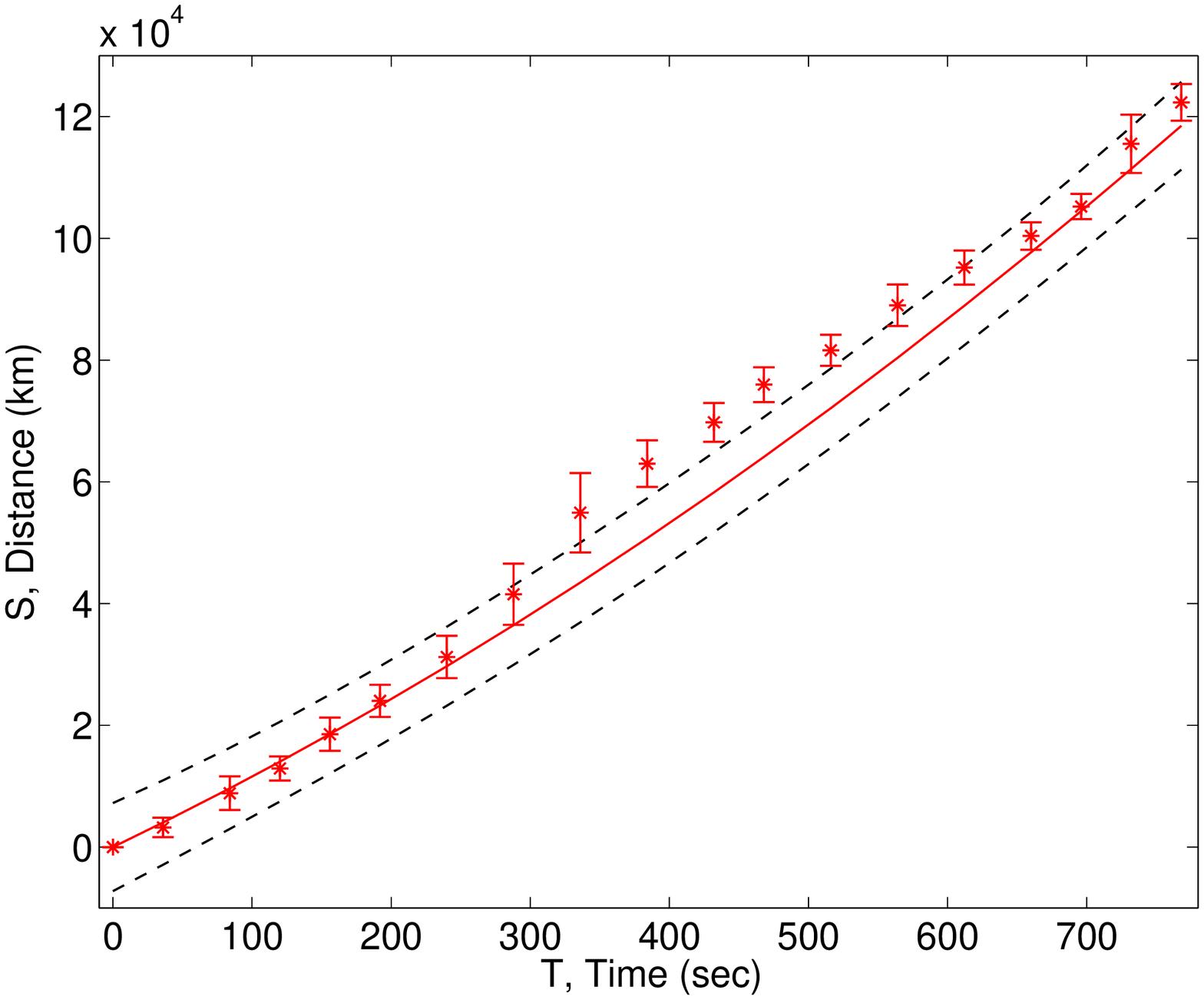}
\includegraphics[width=7cm]{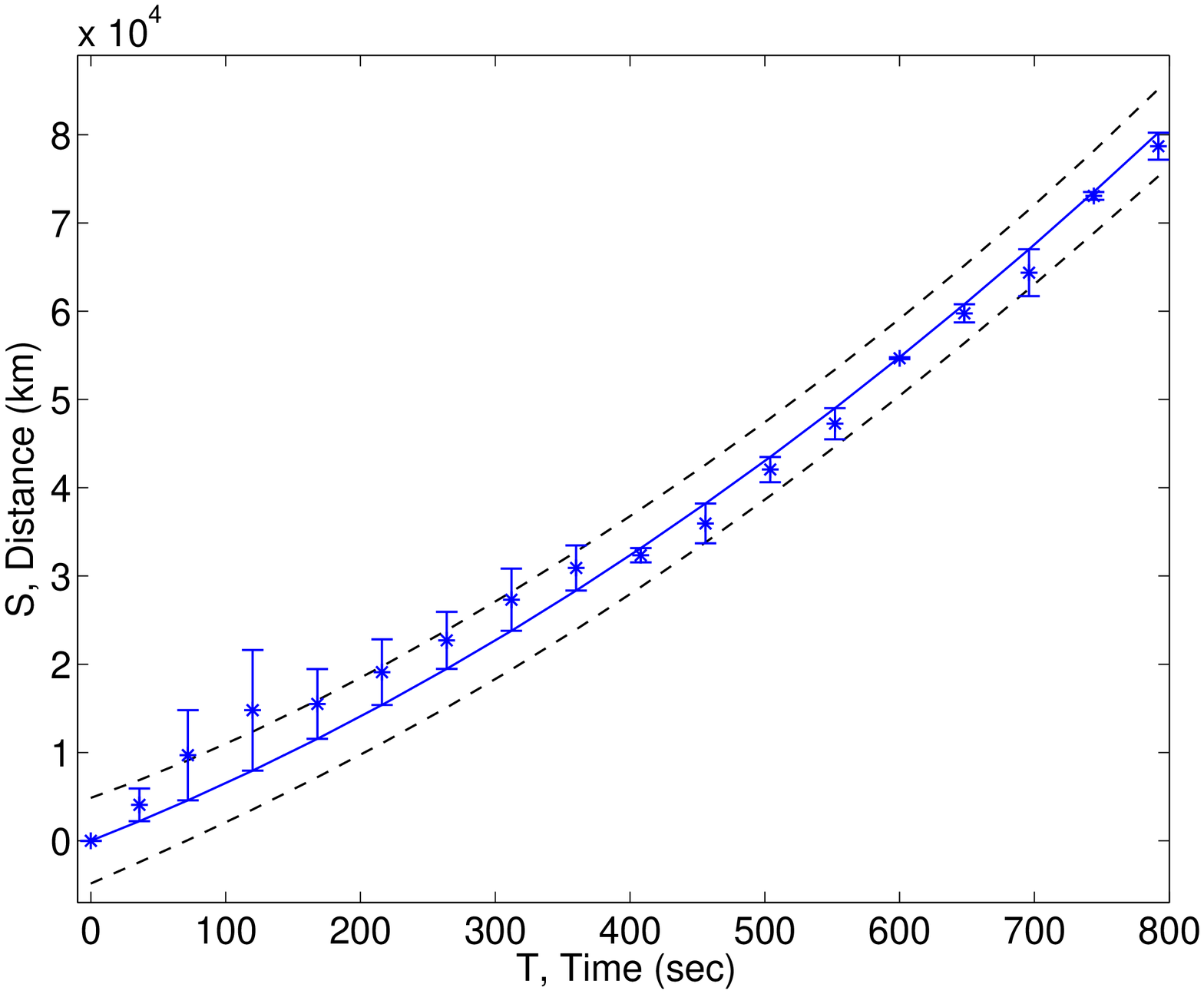}
\end{center}
\caption{Distance vs time for the first (red dots) and second (blue dots) blob of the second sequence. Fitted curves are the trajectories with constant acceleration for the first (red curve, upper panel) and second (blue curve, lower panel) blobs with non-zero initial velocity. }
\end{figure}

\section{Discussion}

Coronal rain is probably connected to coronal heating and thermal instability, therefore it is important to answer the two main questions: how does the coronal rain form, and why is its acceleration lower than gravitational free fall. We used time series of AIA/SDO in the 171 \AA\, and 304 \AA\, spectral lines to answer these questions. We detected the rapid cooling of a coronal loop from 1 MK to 0.05 MK on February 22, 2012. We found that the coronal loop completely disappeared in the 171 \AA\ line and simultaneously appeared in the 304 \AA\ line for $\text{more than}$ one hour. The cooling was accompanied by the formation of coronal rain in the form of falling cold blobs.

\begin{figure*}
\begin{center}
\includegraphics[width=\columnwidth]{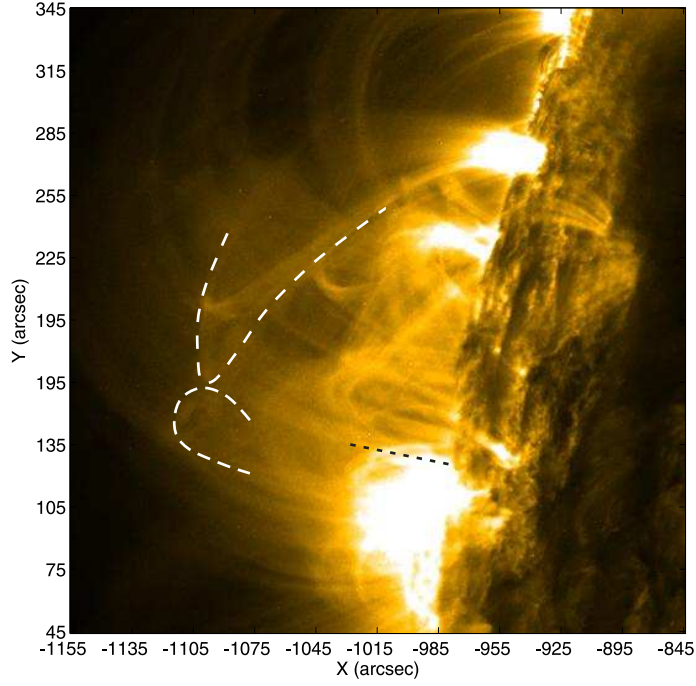}
\end{center}
\caption{Coronal loops that might reconnect and reconstruct the active region are shown as dashed white lines in 171\AA\ above active region AR 11420 at UT 21:00 on February 22, 2012. The dashed black line shows the the trajectory of the coronal rain blobs.}
\end{figure*}

Coronal rain is assumed to be formed by catastrophic cooling when radiative losses locally overcome the heating input (Parker 1953, Field 1965, Antiochos et al. 1999, Schrijver 2001, M\"uller et al. 2003, 2004, 2005), therefore we estimated the energy balance during the cooling. The energy equation for static coronal loops with constant cross-section is (Aschwanden 2004)
\begin{equation}
{{n_e k_B}\over {\gamma-1}}{{\partial T}\over {\partial t}}=E_H-E_R-\nabla F_c,
\label{eq:1}
\end{equation}
where $n_e$ is the electron number density, $T$ is the plasma temperature, $k_B$ is the Boltzmann constant, $\gamma$ is the ratio of specific heats, $E_H$ is the heating rate, $E_R$ is the radiative loss rate, and $F_c$ is the conductive flux.
We estimated the terms on the right-hand side separately. For a typical loop electron number density of $3\times 10^9$ cm$^{-3}$ and a radiative loss function $\Lambda(T)=10^{-22}$ erg cm$^{3}$ s$^{-1}$ corresponding to 1 MK (Rosner et al. 1978), the radiative loss rate is $E_R\approx n^2_e \Lambda(T)= 9\times 10^{-4}$ erg cm$^{-3}$ s$^{-1}$ (Aschwanden 2004). The conductive flux for the loop half-length of 110 Mm (which corresponds to a loop apex height of 70 Mm) is $\nabla F_c\approx 10^{-6}T^{7/2}/L^2 \approx 8\times 10^{-6}$ erg cm$^{-3}$ s$^{-1}$. Therefore, radiative losses seem to be more important than the energy loss by thermal conduction.
The value of the heating function, which needs the loop to be in energy balance, can be estimated either for uniform (Rosner et al. 1978) or nonuniform (Aschwanden and Schrijver, 2002) heating. Using Eq. (4.4) of Rosner et al. (1978), we estimated the heating function as $E_H = 3.37 \times 10^{-4}$ erg cm$^{-3}$ s$^{-1}$. On the other hand, for a loop temperature of 1 MK and an exponential scale height of 110 Mm in the heating function (which corresponds to the loop half-length), the heating rate near the loop top is $\sim 10^{-6}$ erg cm$^{-3}$ s$^{-1}$ (Aschwanden and Schrijver, 2002). Thus, the energy balance is violated as the radiative losses overcome the heat input for both uniform and nonuniform heating. Consequently, this may lead to a catastrophic cooling process. Then, the left-hand term estimates the time interval in which the plasma may cool down from 1 MK to 0.05 MK. The uniform heating function of Rosner et al. (1978) estimates the cooling time as $\sim$ 20 min, while the nonuniform heating gives $\sim$ 10 min. The time interval is smaller than the observed cooling time.
To increase the estimated cooling time, the loop temperature needs to be increased to 1.5 MK, then the heating function of Rosner et al. (1978) estimates the cooling time as $\sim$ 45 min, which is close to the observed time. The recent numerical simulation of Reale et al. (2012) also showed a similar time scale for cooling of post flare loops.

The rapid cooling is immediately followed by the appearance of cool blobs. We traced two different events of coronal rain. The first event occurred below the cooling coronal loop in the form of three blobs during the time interval of UT 21:30-22:15. The first cool blob appeared well below the cooling loop at considerably long distance ($\sim$ 20 Mm). On the other hand, the second blob appeared near the lower border of the coronal loop. But the third blob again appeared at a long distance below the coronal loop. It is an interesting question why two blobs were formed outside the cooling coronal loop. There are three possible answers to this question. First, the blobs are formed inside the coronal loop but remained unnoticed in 304 \AA\, owing to the low mass (and thus low intensity). The blobs then are carried away either by high-speed flows after the thermal instability (M\"uller et al. \cite{Muller2004}) or by reconnection outflows (Murawski et al. \cite{Murawski2011}). Later, the mass (and intensity) of the blobs increased during the fall, and they appeared near the observed heights. In this case, the first and third blobs might already have initial velocities at appearance. If the possible initial velocity of the first blob (estimated as 12 km s$^{-1}$) is used as the mean flow speed, then the time required for the blob to fall from the cooling loop at the observed height is $\sim$ 30 min. This seems to be a reasonable value. Second, magnetic reconnection may reconstruct the active region loops, and the blobs are falling along newly formed loop. The two loops of the active region that were able to reconnect are shown in Fig. 8. Third, the first sequence of blobs occurs in loops that are unrelated to those of the second sequence, but appear to be related as
a result of projection effects. Figure 8 shows a fan of loops emanating from the location from which the first sequence of blobs appeared. \ They reach the previously defined loop system (from which the second sequence originates). These loops seem to match the trajectories of the first sequence of blobs well.

The mean velocities of the first, second, and third blob were estimated to be 50 km s$^{-1}$, 60 km s$^{-1}$ , and 40 km s$^{-1}$ , respectively. The first and third blob followed parabolic trajectories with non-zero acceleration, while the second blob fell down with almost zero acceleration. A polynomial fit of the blob trajectories shows that the estimated accelerations for the first and third blob were different for the motions with and without initial velocity. Without initial velocity, the accelerations of the first and third blob are $\sim$ 120 m s$^{-2}$ and $\sim$ 136 m s$^{-2}$, respectively. With initial velocity, the accelerations of the first and third blob are 92 m s$^{-2}$ and 74 m s$^{-2}$, respectively. In all cases, the acceleration is lower than the gravitational free fall in the solar corona. The first and second blob followed the same paths with finite and zero accelerations, respectively.

Another sequence of coronal rain appeared along inclined paths, as shown by the dotted line in the lower right panel of Fig. 1.
In this case, cool plasma is less clumpy and resembles a flow more than blobs. We were only able to identify two different blobs. The acceleration of the first blob along the inclined loop was estimated to be 130 m s$^{-2}$ with an initial velocity of $V_0$=80 km s$^{-1}$. Considering the inclination of the loop with regard to the vertical, we estimate that this value is lower than the gravitational free-fall acceleration of the solar atmosphere.
On the other hand, the acceleration of the second blob, which follows the same loop after $\sim$ 15 min, is 100 m s$^{-2}$ with an initial velocity of 60 km s$^{-1}$. It is possible that coronal rain blobs follow the first hypothetic blob, falling with stronger acceleration and changing the plasma distribution after its passage. Consequently, the next blobs move with a lower acceleration than the first. This statement needs to be verified by additional observations in the future, but it is possible that the occurrence of this phenomenon depends on the particular parameters of coronal loops.

\section{Conclusion}

Time-series of the 171 \AA\, and 304 \AA\, spectral lines obtained by AIA/SDO show the rapid cooling of a coronal loop from 1 MK to 0.05 MK over one hour. The cooling was accompanied by the appearance of two coronal rain events in the 304 \AA\, line. An energy estimation showed that catastrophic cooling is responsible for the formation of the coronal rain. We observed two different sequences of falling cool blobs. The first sequence included three different blobs. Time-distance analysis revealed that the motion of two different blobs that followed the same path has different characteristics: the first blob moved with non-zero acceleration, the second blob moved without acceleration. Consequently, the acceleration of the blobs does not depend on local loop parameters, but it might depend on the blob mass, as recently suggested by Oliver et al. (2014). Numerical simulations using the model of Oliver et al. (2014) agree well with observations (see Fig.7). The third blob of the first sequence, which followed a different path, moved with non-zero acceleration. However, the acceleration of all blobs was always lower than gravitational free fall, as is usually observed in coronal rain events. All blobs left trails that might be a result of continuous cooling in the blob tails (Fang et al. 2013). The second sequence included two different blobs moving along an inclined path. In this case, the first blob moved with stronger acceleration along the path, while the second blob moved with lower acceleration. We propose that the different accelerations may correspond to different values in the ratio of blob to coronal density: the heavier blobs fall with higher acceleration.

\begin{acknowledgements}
The work was supported by the European FP7-PEOPLE-2010-IRSES-269299 project- SOLSPANET, by the Austrian Fonds zur F\"orderung der Wissenschaftlichen Forschung (projects P26181-N27 and P25640-N27) and by Shota Rustaveli National Science Foundation grant DI/14/6-310/12. R.O. acknowledges the support from the Spanish MINECO and FEDER funds through project AYA201-22846. The work is supported by international Programme "Implications for coronal heating and magnetic fields from coronal rain observations and modeling" of the International Space Science Institute (ISSI), Bern.
\end{acknowledgements}

\end{document}